\documentclass[11pt,a4paper]{article}

\usepackage{fullpage}

\usepackage{ucs}
\usepackage[utf8x]{inputenc}
\usepackage[english]{babel}

\usepackage{amsmath}
\usepackage{amsfonts}
\usepackage{amssymb}

\usepackage{amscd}
\usepackage{graphicx}

\tiny

\author{Julio Guerrero$^1$,  Francisco F. L\'opez-Ruiz$^2$}

\date{\begin{center}
\begin{small}$^1$Departamento de Matem\'atica Aplicada, Universidad de Murcia,
\end{small}\\
\begin{small}Campus de Espinardo, 30100 Murcia, Spain.\end{small}\\
\begin{small}$^2$ Departamento de F\'\i sica Aplicada, Universidad de
C\'adiz,
\end{small}\\
\begin{small}
  Campus de Puerto Real, 11510 Puerto Real, C\'adiz, Spain.
\end{small}\\
\begin{small}juguerre@um.es\ paco.lopezruiz@iaa.es
\end{small}\\                                                  
                                              \end{center}
                                              }

\begin{document}

\title{The Quantum Arnold Transformation and \\ the Ermakov-Pinney equation}

\maketitle

\begin{abstract}

The previously introduced Quantum Arnold Transformation, a unitary operator mapping the 
solutions of the Schr\"odinger equation for time dependent quadratic Hamiltonians into the
solutions  for the free particle, is revised and some interesting
extensions are introduced, providing in particular 
a generalization of the Ermakov-Pinney equation. 

\end{abstract}


%

\section{Introduction}

In \cite{Arnold}, V.I. Arnold introduced a theorem stating that the family of graphs of solutions of any linear second-order
differential equation (LSODE), with arbitrary time dependent coefficients, is locally diffeomorphic to the family of graphs of 
solutions of the simplest one-dimensional equation of motion, i.e., the equation of motion of a free particle. This important 
result allows to establish a direct relationship between the symmetries of an arbitrary classical linear system, whose equation of motion is a LSODE, 
with that of the free particle, which is known to have the largest possible symmetry group, $SL(3,\mathbb{R}$). The diffeomorphism was explicitly given in \cite{Arnold}, namely the Classical Arnold Transformation (CAT), and can be seen as a particular class of more general Lie transformations relating systems defined by second order differential equations with the free particle \cite{Lie}.

In \cite{QAT}, the authors, in collaboration with V. Aldaya and F. Coss\'\i o, introduced the quantum version of the CAT, the 
Quantum Arnold Transformation (henceforth QAT), as a unitary map that relates the Hilbert space of solutions of the time-dependent Schr\"odinger 
equation for a Generalized Caldirola-Kanai oscillator (the quantum version of a classical system
whose equation of motion is a LSODE)
into the corresponding one for the free particle. In the present paper we revise the QAT and extend it in two respects. First, we 
give the explicit transformation connecting quantum physical systems with arbitrary time-dependent quadratic Hamiltonians with the 
free particle system, here called the Gauged QAT. Second and more importantly, we describe the transformation that links any 
LSODE-system to any other LSODE-system in terms of CAT's, giving rise to a generalization of the Ermakov-Pinney equation, and for that
reason we shall call it the Arnold-Ermakov-Pinney transformation, providing also its quantum version. 

Although the QAT is very recent, the CAT is more than thirty years old and, since it was introduced, many authors have used it (even without realizing it) and  have built  something similar to the QAT. Even before that, Lewis and Riesenfeld (1969) 
\cite{Lewis} introduced a technique to obtain solutions of the time-dependent Schr\" odinger equation (TDSE) for a time-dependent quadratic Hamiltonian (TDQH) as eigenfunctions of quadratic invariants. For that purpose they wrote the solutions in terms of auxiliary variables that satisfy the classical equations of motion (something that resembles the CAT). Dodonov \& Man'ko (1979) \cite{Manko} constructed invariant operators for the damped harmonic oscillator and introduced coherent states, using a method similar to that of Lewis and Riesenfeld.
Jackiw (1980) \cite{Jackiw} gave (implicitly) the quantum transformation from the harmonic oscillator (even with a $1/x^2$ term)
to the free particle when studying the symmetries of the magnetic monopole.
Junker \& Inomata (1985) \cite{Junker} gave the transformation of the propagator,
in a  {path integral} approach, for a TDQH, into 
the free one (the equivalent of the QAT, but in terms of propagators).
Takagi (1990) \cite{Takagi} gave the quantum transformation from
 the harmonic oscillator to the free particle, interpreted as the change to
{comoving} coordinates.
Bluman \& Shtelen (1996) \cite{Bluman} gave the {(non-local)} transformation of the TDSE for
a TDQH plus a {non-linear} term into the free particle
one, in the context of transformations of PDEs. Kagan et al. \cite{Kagan} and independently  Castin \& Dum  \cite{Castin} 
(1996)  introduced
a scaling transformation in the Gross-Pitaevskii equation describing Bose-Einstein Condensates (BEC) which is related to the
QAT. Suslov et al. (2010) \cite{Suslov}  computed the propagator for a TDQH using the classical equations.

To the best of the authors' knowledge, the Quantum Arnold Transformation can be not only very useful to perform certain analytic calculations, but key to unify different concepts scattered through the mentioned literature in a simplified framework. 

The paper is organized as follows. First, in Section \ref{sec:CAT} we revise the Classical Arnold Transformation. After that we go on with the Quantum Arnold Transformation in Section \ref{sec:QAT}. Finally two extensions are described in Section \ref{sec:ext}: the Gauged QAT and the Arnold-Ermakov-Pinney transformation.

\section[Arnold Transformation]{The Classical Arnold Transformation}
\label{sec:CAT}

To set up the framework in which the classical Arnold transformation acts, let us briefly remind the subject of Lie symmetries of
ordinary differential equations.

\subsection{Lie transformations}\label{sec:Lie}

A Lie symmetry of an ordinary differential equation (ODE) is a coordinate transformation that sends solutions into solutions.
The problem of determining the Lie symmetries of an ODE is rather old,
and S. Lie gave the main results at the end of the nineteenth century \cite{Lie}. One
of these results was that a second order differential equation (SODE)
$y'' = F(x,y,y')$
has the maximal number of Lie symmetries ($SL(3,\mathbb{R})$) if it can 
be transformed to the {free equation} by a {point transformation}:
\begin{equation}
y'' = F(x,y,y')  
\stackrel{\tiny\begin{array}{c}
         {\tiny  \tilde{x}=\tilde{x}(x,y)}\\
          { \tilde{y}=\tilde{y}(x,y)}
          \end{array}}{\Longrightarrow}  
\tilde{y}''=0\,. \label{Lie}
\end{equation}
This {linearization} is possible if the ODE is of the form:
\begin{equation}
 y''=E_3(x,y)(y')^3 + E_2(x,y)(y')^2+ E_1(x,y)y'+E_0(x,y)\,, \label{EDO}
\end{equation}
with $E_i(x,y)$ satisfying some \textit{integrability conditions} (see, for instance, \cite{Mahomed-Qadir,Aminova}).

There is a nice geometric interpretation of this condition in terms of projective geometry. 
The non-linear SODE (\ref{EDO}) is obtained by projection from  the {geodesic equations} in a 2-dim Riemannian 
manifold. The coefficients $E_i(x,y)$ are in one-to-one correspondence with  {Thomas projective parameters} $\Pi$, and 
the  integrability conditions that they satisfy are the conditions for the  Riemann tensor to be zero
(see \cite{Mahomed-Qadir,Aminova}).

V.I. Arnold 
named this process \textit{rectification} or  \textit{straightening} of the trajectories, and studied the case of linear {SODE} 
(LSODE), giving explicitly the point transformation for this case \cite{Arnold}. In the next subsection we
describe it in detail.

 
\subsection{Classical Arnold transformation}

A General {L}inear {S}econd {O}rder
{D}ifferential {E}quation ({LSODE}) is given by the differential equation:
\begin{equation}
 {\ddot{x}+\dot{f}\dot{x} +\omega^{2}x=\Lambda}\,,\label{LSODE}
\end{equation}
where $f, \omega$ and $\Lambda$ are functions of $t$.
The Classical Arnold Transformation (CAT) is a point transformation that is a local diffeomorphism: 
\begin{equation}
\begin{array}{rccl}
{A} :& \mathbb{R}\times T&\rightarrow &\mathbb{R}\times{\cal T}\\
&(x,t)&\mapsto &(\kappa,\tau)
\end{array}
\,\,:
\left\{\begin{array}{rl}
\tau&=\frac{u_{1}(t)}{u_{2}(t)}=\int_{t_0}^t\frac{W(t')}{u_2(t')^2}dt'\\
\kappa&=\frac{x - u_p(t)}{u_{2}(t)}
\end{array}\right.\,,
\end{equation}
where $T$ and ${\cal T}$ are, in general, open intervals containing $t_0$ and $0$, respectively,
$u_1$ and $u_2$ are independent solutions of the homogeneous LSODE satisfying the 
canonicity conditions:
\begin{equation}
u_1(t_0)=0=u_2'(t_0)\,,\qquad u_1'(t_0)=1=u_2(t_0)\,,\label{canonicity}
\end{equation}
$u_p$ is a particular solution of the inhomogeneous LSODE satisfying $u_p(t_0)=u_p'(t_0)=0$, 
and $W(t)=\dot{u}_1u_2-u_1\dot{u}_2=e^{-f}$ is the { Wronskian} of the two solutions.  
Here $t_0$ is an arbitrary time, 
conveniently chosen to be $t_0=0$ (see \cite{QAT} for details).

The CAT transforms the original LSODE (\ref{LSODE}) into that of the free particle, up to a factor:
\begin{equation}
\ddot{x}+\dot{f}\dot{x} +\omega^{2}x=\Lambda  \quad
  \stackrel{{A}}{\longrightarrow}\quad \frac{W}{u_2^3}\,\, \ddot \kappa =0\,.
\label{eq:difeomorfismo no homogeneo completo}
\end{equation}

The presence of this factor
implies that  patches of trajectories of (\ref{LSODE}) 
are transformed into patches of straight (free) trajectories. In fact, an arbitrary trajectory solution of (\ref{LSODE}) can be written as
 $x(t)=A u_1(t) + Bu_2(t)+u_p(t)$, and the CAT sends it to  $\kappa(\tau)=A\tau + B$. While
$t$ varies in the interval $T$ defined by two consecutive zeros of $u_2(t)$ (containing $t_0$), $\tau$ varies in the range of the map 
defined by $\frac{u_{1}(t)}{u_{2}(t)}$. In the case in which $u_2(t)$ has one zero, $T$ is (left- or right-) unbounded, and, if it has no zeros,  $T$ is $\mathbb{R}$. 

Even though the CAT is a local diffeomorphism, it can be defined for an arbitrary time $t_0$. Thus 
different CATs can be defined for different times $t_0$ and cover in this way a complete trajectory
of (\ref{LSODE}). We shall show with the example of the harmonic oscillator how this can be done.
 
\subsubsection{The example of the harmonic oscillator}

The harmonic oscillator (HO) is the best  example to understand how the CAT works. For this case, and considering 
$\Lambda=0$, the two solutions  are $u_1(t)=\frac{1}{\omega}\sin(\omega t)$ 
and $u_2(t)=\cos(\omega t)$. The open interval $T$ defined by two consecutive zeros of $u_2(t)$, and containing $t_0=0$, is $(-\frac{\pi}{2\omega},\frac{\pi}{2\omega})$, and the  CAT $A$ and its inverse $A^{-1}$ are then written as:
%
%

 \begin{equation}
  A:\,\left\{ \begin{array}{cccc} 
 \kappa&=\frac{x}{u_2(t)} = \frac{x}{\cos(\omega t)}\,,& 
 \tau&=\frac{u_1(t)}{u_2(t)}= \frac{1}{\omega}\tan(\omega t)
  \end{array}\right.\label{CAT-HO} 
 \end{equation}
 \begin{equation}
 A^{-1}:\,\left\{ \begin{array}{cccc}                
 x&=\cos(\arctan(\omega \tau))\kappa\,, &
 &t=\frac{1}{\omega}\arctan(\omega t)\,.\\
  &=\frac{\kappa}{\sqrt{1+\omega^2 \tau^2}}& \hbox{\ } &
 \end{array}\right.\label{CAT-HO-inv}\
 \end{equation}

\begin{figure}[ht]
\centering
 \includegraphics[width=7cm]{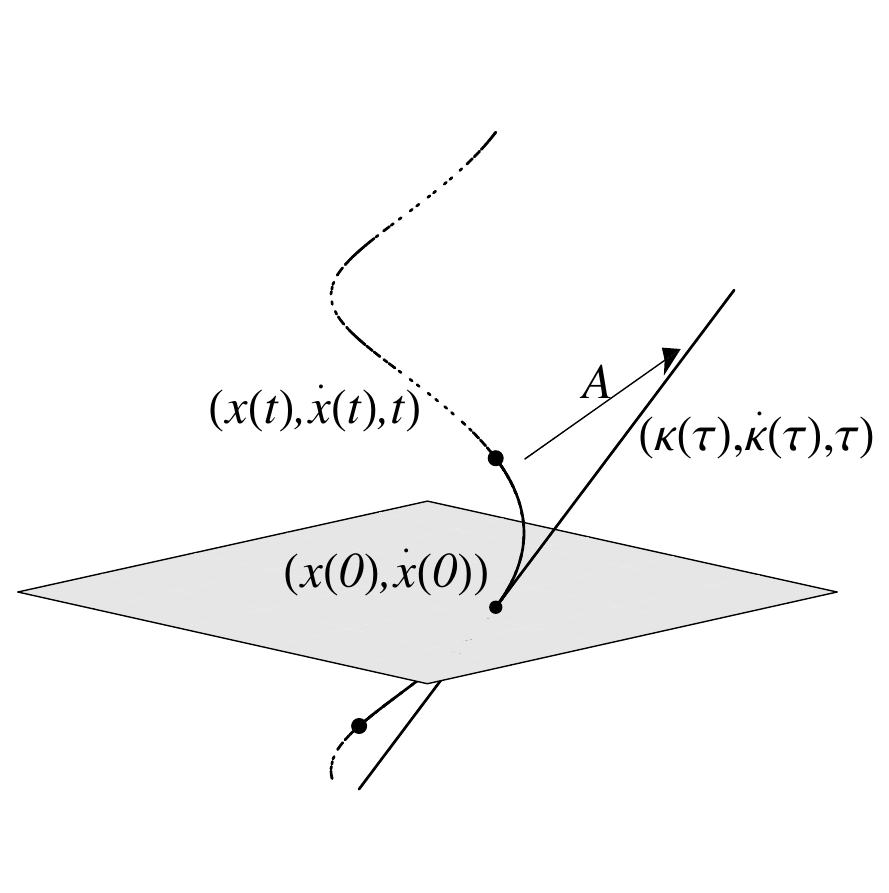}
\caption{Depiction of the CAT for the harmonic oscillator (adapted from \cite{HarmonicStates}).} 
\label{figura-CAT}
\end{figure}

In this case $\tau\in\mathbb{R}$. 
Pictorially, the CAT for the HO can be represented as in Figure~\ref{figura-CAT}, where velocities have also been included in the
graphic for clarity. 
Here $A$  maps the solid part of the helix (half a period of a harmonic
oscillator trajectory) into the whole line (a free particle trajectory). The
horizontal plane represents the space of all possible initial conditions
at $t=0=\tau$.  Note that both trajectories  are tangent when projected onto this plane, due to
the conditions  (\ref{canonicity}).
See \cite{HarmonicStates} for more details in this case.
For the CAT to map other patches of the HO trajectories into the free particle trajectories,
different branches of the $\arctan$ function in the inverse CAT (\ref{CAT-HO-inv}) should be used (and a different $t_0\neq 0$ 
for the CAT).  
For each integer $k$,  let us take $T_k=((k-\frac{1}{2})\frac{\pi}{\omega},(k+\frac{1}{2})\frac{\pi}{\omega})$  and $t_k=k\frac{\pi}{\omega}$. The solutions verifying 
conditions (\ref{canonicity}) at $t_k$ are $u_i^{(k)}(t)=(-1)^k u_i(t)=u_i(t-t_k)\,,\,i=1,2$.
Define a pair of CAT and inverse CAT  from $\mathbb{R}\times T_k$ into $\mathbb{R}^2$ of the form:
$A_{(k)}(x,t)=(\frac{x}{u_2^{(k)}(t)},\frac{u_1^{(k)}(t)}{u_2^{(k)}(t)})=(\kappa,\tau)$ and 
$A_{(k)}^{-1}(\kappa,\tau)= (x,t)$, where  the $k$-th branch of the $\arctan$ function has been used in  $A_{(k)}^{-1}$. 
An \textit{unfolded} version of the CAT, $\tilde{A}$, can be built by joining all the patches $A_{(k)}$, defining an application
that maps a complete trajectory $x(t)$ of the harmonic oscillator into a  trajectory $\kappa(\tau)$ of the free particle. $\tilde{A}$ is periodic
on $t$ with period $\frac{\pi}{\omega}$, although discontinuous.

Other simple examples where this construction can be done are the damped particle and the damped harmonic oscillator,
see \cite{UnfoldedQAT} for details.

%
%
%
%
%
%
%
%
%
%
%
%
%
%
%
%


\section{The Quantum Arnold Transformation}\label{sec:QAT}

 
An arbitrary LSODE system (\ref{LSODE}) can be derived from the Lagrangian (we shall take $\Lambda=0$
for simplicity but the whole formalism can be developed with $\Lambda\neq 0$, see \cite{QAT}):
\begin{equation}
L=\frac{1}{2}m e^{f}\left(\dot{x}^2-\omega^2 x^2\right)\,,
\end{equation}
and from this the Hamiltonian 
\begin{equation}
H=\frac{p^{2}}{2m}e^{-f}+\frac{1}{2}m\omega^{2}x^{2} 
e^{f}\,
\label{CKH}
\end{equation}
is derived, which is known as the Generalized Caldirola-Kanai (GCK) Hamiltonian  for a damped oscillator (see \cite{QAT} and references therein). 
The case in which $\dot{f}=\gamma$ and $\omega$ are
constants corresponds to the original Caldirola-Kanai Hamiltonian for a damped harmonic oscillator 
\cite{Caldirola,Kanai}, and whose 
corresponding Lagrangian was given for the first time by Bateman \cite{Bateman}, after a constraint to the
Bateman dual system is imposed \cite{Bateman-nuestro}.
Canonical quantization of the GCK Hamiltonian leads to the time-dependent Schr\"odinger equation:
\begin{equation}
i\hbar\frac{\partial \phi}{\partial t}=\hat{H}\phi=-\frac{\hbar^{2}}{2m}e^{-f}
\frac{\partial^{2} \phi}{\partial x^{2}}+\frac{1}{2}m\omega^{2}x^{2}
e^{f}\phi \,.
\label{CKeq}
\end{equation}

The CAT $A$ is a local (in time) diffeomorphism between the space of solutions of the LSODE system (\ref{LSODE}) and the space of 
solutions of the free
particle. We would like to extend it to a unitary transformation $\hat{A}$, the Quantum Arnold Transformation (QAT), between the 
Hilbert space of solutions $\phi(x,t)$ of the time-dependent Schr\"odinger equation for the GCK oscillator (\ref{CKeq}) at time $t$,
${\mathcal H_t}$, into the Hilbert space of solutions $\varphi(\kappa,\tau)$ of the time-dependent Schr\"odinger equation for the Galilean
free particle
\begin{equation}
 i\hbar\frac{\partial \varphi}{\partial \tau}=-\frac{\hbar^{2}}{2m}
\frac{\partial^{2} \varphi}{\partial \kappa^{2}} \,,\label{free-eq}
\end{equation}
at time $\tau$, ${\mathcal H^G_\tau}$. The desired extension is 
given by: 
%
\begin{align}
  \hat {A}:& \quad {\mathcal H_t} \;\; \longrightarrow
\quad {\mathcal H^G_\tau} \nonumber\\
    & {\phi}(x,t)  \longmapsto \;
{\varphi}(\kappa,\tau) = 
        \hat A \left( {\phi}(x,t) \right) \\
     & \qquad \qquad = 
       A^* \left( \sqrt{u_{2}(t)}\,e^{-\frac{i}{2}\frac{m}{\hbar}
                 \frac{1}{W(t)}\frac{\dot{u}_{2}(t)}{u_{2}(t)} {x}^{2}}
{\phi}(x,t) \right) \,.   \nonumber
\end{align}
%
Here $A^*$ is the pullback of the CAT $A$, acting on functions (i.e. $A^*(f(x,t))=f(A^{-1}(\kappa,\tau))$).
The QAT can be diagrammatically represented as:
\begin{equation}
\begin{CD} 
{\mathcal H^G_\tau} @<\hat{A}<<
{\mathcal H_t}\\ 
@A{\hat U_G(\tau)}AA @AA\hat U(t)A\\ 
\mathcal H^G_0 \equiv \mathcal H @>>\hat{1}> \mathcal H \equiv \mathcal H_0 
\end{CD}\label{diagrama-QAT}
\end{equation}
where $\mathcal{H}_0\equiv\mathcal{H}^G_0\equiv \mathcal{H}$ is the common Hilbert space of solutions of the Schr\"odinger
equation for both systems at $t=\tau=0$ (we shall take, for simplicity, $t_0=0$, as before), $U(t)$ is the unitary time-evolution operator for the  GCK oscillator and 
${\hat U_G(\tau)}$ is the corresponding one for the Galilean free particle. The map  at the
bottom of the diagram is the identity due to conditions (\ref{canonicity}); otherwise a non-trivial unitary
transformation appears (see \cite{QAT}).

From the commutative diagram, it is clear that $\hat A$ is unitary given the unitarity of the evolution operators. However, it can also be checked explicitly that the scalar product of two states in $\mathcal H^G_\tau$ at a given time $\tau$ is the same than that of the transformed states by $\hat A$ in $\mathcal H_t$ at the corresponding time $t$: 

\begin{align}
\langle \varphi_1 , \varphi_2 \rangle_{\mathcal H^G_\tau} 
&= 
\int_{-\infty}^{+\infty} {\rm d} \kappa \,\varphi_1(\kappa,\tau)^*\varphi_2(\kappa,\tau) \nonumber
\\
&=
\int_{-\infty}^{+\infty} \frac{{\rm d}x}{u_2(t)} \left( \sqrt{u_{2}(t)}\,e^{\frac{i}{2}\frac{m}{\hbar}
                 \frac{1}{W(t)}\frac{\dot{u}_{2}(t)}{u_{2}(t)} {x}^{2}}
{\phi}_1(x,t)^* \right)  \label{unitarityQAT}
\\
& \qquad \qquad \quad \times\left( \sqrt{u_{2}(t)}\,e^{-\frac{i}{2}\frac{m}{\hbar}
                 \frac{1}{W(t)}\frac{\dot{u}_{2}(t)}{u_{2}(t)} {x}^{2}}
{\phi}_2(x,t) \right) \nonumber
\\
&= 
\int_{-\infty}^{+\infty} {\rm d} x \,\phi_1(x,t)^*\phi_2(x,t)
=
\langle \phi_1 , \phi_2 \rangle_{\mathcal H_t} \,, \nonumber
\end{align}
where $\tau$, $\kappa$, ${\rm d}\kappa$ and the integration limits have been transformed according to the CAT. 

The QAT inherits from the CAT the local character in time, in the sense that it is valid only for $t\in T$ and $\tau\in \mathcal{T}$, although it can be defined for an arbitrary initial
time $t_0$. To extend the QAT beyond $T$, we can proceed as in the classical case for the harmonic oscillator, considering
the different branches of the inverse function of $\tau(t)$, defining an \textit{unfolded} QAT, $\hat{\tilde{A}}$. 

It should
be stressed that if in the different branches of the \textit{unfolded} CAT proper solutions verifying (\ref{canonicity}) are not used, 
changes in signs
can appear, which result in changes in phases in the different branches of the \textit{unfolded} QAT. This phenomenon is related to the Maslov correction
(see for instance \cite{Horvathy}). In fact, it can be checked that, for the case of the harmonic oscillator
previously considered, $\hat{A}_{(k)}(\phi(x,t))=e^{ik\frac{\pi}{2}}\hat{A}(\phi((-1)^k x,t))$.

\subsection{Symmetries of the quantum GCK oscillator}

By the results of Sec. \ref{sec:Lie}, any classical LSODE system has the maximal number of
Lie symmetries, the ones of the free particle ($SL(3,\mathbb{R})$). The vector fields generating these 
symmetries can be computed by means of the CAT from the symmetries of the free particle. Even though 
the CAT is a local diffeomorphism, these vector fields are defined for all times \cite{GonzalezLopez} 
(the reason for this is that the only function appearing in the denominator of the vector fields is the 
Wronskian, which is never vanishing). 

The quantum symmetries 
of the free Schr\"odinger equation (\ref{free-eq}) are smaller than $SL(3,\mathbb{R})$, only the Schr\"odinger group
preserves\footnote{Symmetries of the Schr\"odinger equation are the quantum analogue of Noether symmetries, i.e. those
preserving the Lagrangian.} (\ref{free-eq}) \cite{Niederer}. The Schr\"odinger group is a semidirect product
of the (centrally extended) Galilei group by the group $SL(2,\mathbb{R})$ . It is generated by the 
basic operators:
\begin{equation}
 \hat \kappa = \kappa + \frac{i \hbar}{m} \tau
\frac{\partial}{\partial \kappa}\,,
\qquad
\hat \pi = -i \hbar \frac{\partial}{\partial \kappa}\,,\label{basic-free}
\end{equation}
in addition to their quadratic combinations (representing scale and non-relativistic 
\textit{conformal} transformations).

The basic operators, together with their quadratic combinations, are constant of motion operators\footnote{They verify that 
$\frac{\partial\cdot }{\partial t} +\frac{i}{\hbar}[\hat{H}^G,\cdot]=0$, where 
$\hat{H}^G=-\frac{\hbar^{2}}{2m} \frac{\partial^{2}}{\partial \kappa^{2}}$ is the Hamiltonian 
operator for the free Galilean particle.}, they are infinitesimal generators of symmetries of the 
free Schr\"odinger equation (\ref{free-eq}), and preserve the Hilbert space $\mathcal H^G_\tau$ 
of solutions (\ref{free-eq}).

We can import the symmetries acting on the free particle Hilbert space 
${\mathcal H}^G_\tau$ (the Schr\"odinger group) into the GCK Hilbert space ${\mathcal H_t}$
by means of the commutative diagram (\ref{diagrama-QAT}). The infinitesimal generators of these
symmetries are:
\begin{equation}
\hat P = -i \hbar u_2 \frac{\partial}{\partial x} - m x \frac{\dot{u}_2}{W}\,,\qquad
\hat X = \frac{\dot{u}_1}{W}x + \frac{i\hbar}{m} u_1\frac{\partial}{\partial x}\,,\label{basic-GCK}
\end{equation}
and their quadratic combinations (see \cite{QAT} for the complete list of operators).
All these operators expand a representation of the Schr\"odinger group, are constant of
motion operators for (\ref{CKeq}) and preserve the Hilbert space $\mathcal H_t$.

Despite of the local character of the QAT, and similarly to the classical case, these operators are 
well defined for all $t\in \mathbb{R}$, since the Wronskian $W$ never vanishes.

The Hamiltonian for the GKC harmonic oscillator can be expressed in terms of the basic, conserved 
operators:
\begin{equation}
\hat{H} = \alpha(t)\frac{\hat P^2}{2m}+\frac{1}{2} m \beta(t) \hat X^2
+ \delta(t) \frac{1}{2}(\hat X\hat P+\hat P\hat X)\,,
\end{equation}
where the time-dependent functions $\alpha(t),\beta(t)$ and $\delta(t)$ are given by:
\begin{equation}
 \alpha(t)=\frac{u_1^2 \omega^2+u_1'^2}{W}\,,\quad
\beta(t)= \frac{u_2^2 \omega^2+u_2'^2}{W}\,,\quad
\delta(t)= \frac{u_1 u_2 \omega^2+u_1'u_2'}{W}\,.
\end{equation}

Again, since the Wronskian never vanishes and the basic operators $\hat X$ and $\hat P$ are also
well-defined for all times, this construction remains valid for all times.

Note that, since the basic operators $\hat X$ and $\hat P$ are conserved, $\hat{H}$
is not a conserved operator (unless $\omega=\omega_0$ is a constant and $\gamma=0$), 
and therefore an eigenvalue equation for it makes no sense 
(i.e., there is no time-independent Schr\"odinger equation).

Instead, any linear combination of $\hat P^2$, $\hat X^2$ and $\hat{XP}$ can be used to
define an eigenvalue problem, and the corresponding eigenvectors would constitute a basis for the
solution of the Generalized Caldirola-Kanai 
Schr\"odinger equation (see \cite{QAT} and references therein).

\subsection{Applications of the QAT}\label{sec:Appl}

{}From the commutative diagram (\ref{diagrama-QAT}) and from (\ref{unitarityQAT}) it is clear that the QAT  is a unitary operator, and this has interesting
and far-reaching consequences. Among them we can mention the possibility of importing operators (symmetries) from one system to the other, 
importing wave functions, scalar product, computing the time evolution operator, etc. We shall not discuss them here, referring the reader to 
\cite{QAT,HarmonicStates,CQAT} for details.

The QAT  have many more and very interesting applications and generalizations. One of them is 
 the application of
the QAT to {density matrices}, transforming the {Quantum Liouville equation} of an arbitrary time-dependent quadratic
Hamiltonian into the free one \cite{HarmonicStates}, with the possibility of
extending it and to {Wigner functions} and other quasi-probability distributions,
or even the extension of the QAT to Lindblad type equations, in order to study dissipation and decoherence under the QAT point of view. 
It can also be applied  to Bose-Einstein Condensates, to transform the time-dependent 
potential (oscillator traps with time-dependent frequencies) into a time-independent potential.

The QAT
can also be relevant in  quantum inflationary Cosmological models, where quantum fluctuations of the scalar field are governed,
under some assumptions, by a LSODE with a dissipative term.

Among the possible generalizations of the QAT, we mention the relativistic case,  as the quantum version of 
a generalized CAT for geodesic equations in a fixed background and with external forces, or
as the quantum version of \textit{geodesic mappings}, that transform geodesics of a metric into geodesics of a different
metric (Beltrami Theorem). Other possible generalizations refer to non-linear potentials (the Quantum Lie Transformation), to second order Riccati equations (see for instance \cite{s_o_RiccEq}), or even to non-local potentials like in the Gross-Pitaevskii equation.

\section{Extensions of the QAT}
\label{sec:ext}

In this section we shall introduce the main results of the paper: 
two
immediate, although non-trivial, extensions of the QAT. The first one is to allow
for more general Schr\"odinger equations than the GKC one. The second one is the transformation relating two LSODE-systems, denoted as the Arnold-Ermakov-Pinney transformation, obtained as the composition of the  CAT corresponding to the first LSODE-system with the inverse of the CAT corresponding to the second LSODE-system, as well as the corresponding quantum version.

\subsection{Gauged Quantum Arnold Transformation}

The QAT, as it has been designed, transforms the solutions of the Schr\"odinger equation of the GKC  oscillator 
into that of the free particle. 
It cannot transform, as it stands, the solutions of the Schr\"odinger equation associated with an arbitrary time-dependent quadratic 
Hamiltonian.

 The reason is the choice 
of solutions $u_1(t),u_2(t)$ satisfying the canonicity condition (\ref{canonicity}) and the fact
that the classical counterpart, the CAT, transforms the LSODE (\ref{LSODE}) into that of the
free particle.

Suppose that the quantum Hamiltonian is an arbitrary, time-dependent self-adjoint quadratic Hamiltonian:
\begin{equation}
 \hat H=-\frac{\hbar^2}{2m} \mu(t)\frac{\partial^2\ }{\partial x^2} -
i\Gamma(t)\hbar\left(x \frac{\partial\ }{\partial x}+\frac{1}{2}\right)+
\frac{1}{2}m \nu(t) x^2 \,,\label{Harbitrario}
\end{equation}
where $\Gamma(t)$ and $\nu(t)$ are real and $\mu(t)>0$ (to have a positive-definite kinetic energy).

Then the transformation that maps solutions of the Schr\"odinger equation for this Hamiltonian into that of
the free particle is given by:

\begin{equation}
 \varphi(\kappa,\tau) = 
        \hat G \left( {\phi}(x,t) \right) = 
       A^* \left( \sqrt{u_{2}(t)}\,e^{-\frac{i}{2}\frac{m}{\hbar}
                 \frac{1}{W(t)}\left(\frac{\dot{u}_{2}(t)}{u_{2}(t)}-
                 \Gamma(t)\right) {x}^{2}}
{\phi}(x,t) \right)\,,\label{gaugedQAT}
\end{equation}
where $u_1(t)$ and $u_2(t)$ are canonical solutions (in the sense of (\ref{canonicity})) 
verifying the LSODE (\ref{LSODE}) with $\dot{f}=-\frac{\dot{\mu}}{\mu}$ and 
$\omega^2= \mu \nu+\Gamma( \frac{\dot{\mu}}{\mu}-\Gamma)-\dot\Gamma$. The Wronskian of the two solutions
is $W(t)=\mu(t)$.

Note that the transformation (\ref{gaugedQAT}) is an ordinary QAT times an extra \emph{gauge}
term $e^{\frac{i}{2}\frac{m}{\hbar} \frac{\Gamma(t)}{W(t)}x^2}$:
\begin{equation}
 \hat G = A^*(e^{\frac{i}{2}\frac{m}{\hbar} \frac{\Gamma(t)}{W(t)}x^2}) \hat A\,.
\end{equation}
For this reason we shall denote this transformation Gauged QAT (GQAT).
In the case where $\Gamma(t)=0$, the GQAT turns into an ordinary QAT.

Note that $M(t)=\frac{m}{\mu(t)}$ can be interpreted as a time-dependent mass, and therefore the 
Hamiltonian can be written as 

\begin{equation}
 \hat{H}=-\frac{\hbar^2}{2 M(t)} \frac{\partial^2\ }{\partial x^2} -
i\Gamma(t)\hbar\left(x \frac{\partial\ }{\partial x}+\frac{1}{2}\right)+
\frac{1}{2}M(t) \Omega(t)^2 x^2\,,\label{variable-mass}
\end{equation} 
where $\Omega(t)^2=\mu(t)\nu(t)$, and the \textit{classical} damping coefficient 
and frequency 
are written now as $\dot{f}=\frac{\dot{M}}{M}$ and $\omega^2= \Omega^2-\Gamma( \frac{\dot{M}}{M}+
\Gamma)-\dot\Gamma$, respectively.

\subsection{The Arnold-Ermakov-Pinney transformation}   

Although the CAT  relates any LSODE system with the free particle, the dynamics of both systems
can be very different\footnote{Consider, for example, the harmonic oscillator with a bounded, periodic motion as compared
with the free particle, with an unbounded non-periodic motion.}. Thus, it could be interesting to relate directly two arbitrary
LSODE systems with similar behaviour, and this can be achieved by
composing a CAT and an inverse CAT. 

More precisely, let $A_1$ and $A_2$ denote
the CATs relating the LSODE-system 1 and LSODE-system 2  to the free particle, respectively, then 
$E=A_1^{-1}A_2$ relates LSODE-system 2 to LSODE-system 1. $E$ can be written as:
\begin{align} 
 E: &\,\mathbb{R}\times T_2\rightarrow \mathbb{R}\times T_1 \nonumber \\
&(x_2,t_2)\mapsto (x_1,t_1)=E(x_2,t_2)\,.
\end{align}

The explicit form of the transformation can be easily computed by composing the two CATs, resulting in:
\begin{equation}
x_1=\frac{x_2}{b(t_2)}\qquad W_1(t_1)dt_1=\frac{W_2(t_2)}{b(t_2)^2}dt_2\,,
\label{Arnold-Ermakov-Pinney}
\end{equation}
where $b(t_2)=\frac{u^{(2)}_2(t_2)}{u^{(1)}_2(t_1)}$ satisfies the non-linear SODE:
\begin{equation}
 \ddot b + \dot f_2 \dot b + \omega_2 b = \frac{W_2^2}{W_1^2}\frac{1}{b^3}
            \left[\omega_1^2+\dot f_1 \frac{\dot u^{(1)}_2}{u^{(1)}_2}(1-b^2\frac{W_1}{W_2})\right]\,,
\end{equation}
and where $u^{(j)}_i$ refers to the $i$-th particular solution for system $j$; $W_j$, $\dot f_j$ and $\omega_j$ stand for the 
Wronskian and the LSODE coefficients for system $j$; and the dot means derivation with respect to the corresponding time function.

{}For the particular case where LSODE-system 1 is a harmonic oscillator ($\omega_1(t_1)\equiv \omega_0$ and $\dot f_1=0$), this
expression simplifies to:
\begin{equation}
 \ddot b + \dot f_2 \dot b + \omega_2 b = \frac{W_2^2}{b^3}\omega_0^2\,,\label{generalized-Ermakov-Pinney}
\end{equation}
resulting in a generalization of the Ermakov-Pinney equation. For $\dot f_2=0$ the Ermakov-Pinney  equation 
(also known as Milne-Pinney) is 
recovered \cite{Ermakov,Milne,Pinney}:
\begin{equation}
 \ddot b + \omega_2 b = \frac{\omega_0^2}{b^3}\,,\label{Ermakov-Pinney}
\end{equation}
representing a harmonic oscillator of frequency $\omega_2$ with an extra inverse squared potential $\frac{\omega_0^2}{x^2}$.
For $\omega_0=0$, the Arnold-Ermakov-Pinney transformation reduces to the ordinary CAT, i.e.  $E=A$.

The Ermakov-Pinney equation is related to the Ermakov invariant \cite{Ermakov,Lewis}
and appears in many branches of physics, such as Cosmology \cite{cosmology}, BEC \cite{Dieter}, etc.
Its generalization to higher dimensions, know as Ermakov pairs (or system), appears 
in BEC \cite{Kagan,Castin} and what is known as Kepler-Ermakov systems \cite{Kepler-Ermakov}.

The Ermakov-Pinney equation entails a kind of nonlinear superposition principle, in the sense
that its solutions can be written in terms of the solutions $y_1(t),y_2(t)$ of the corresponding  linear equation
(with $\omega_0=0$):
\begin{equation}
b(t)^2=c_1 y_1(t)^2 + c_2 y_2(t)^2 + 2c_3 y_1(t)y_2(t)\,,\quad c_1c_2-c_3^2=\omega_0^2\,.
\label{oscilador-Pinney}
\end{equation}

The other way round, the general solution $y(t)$ of the linear equation can be writen in terms of a 
paricular solution $\rho(t)$ of the Ermakov-Pinney equation (\ref{Ermakov-Pinney}) as:
\begin{equation} 
y(t)=c_1\rho(t)\cos(\omega_0\, \theta(t) +c_2)\,,\label{Pinney-oscilador}
\end{equation}
where $c_1,c_2$ are arbitrary constantss  and
$\theta(t)=\int^t \rho^{-2}dt'$. Note that this equation is just (\ref{Arnold-Ermakov-Pinney})
for $W_1=W_2=1$, $t_1=\theta(t_2)$, $\rho=b$, $x_2=y(t_2)$ and $x_1=y(t_2)/b(t_2)=c_1\cos(\omega_0 t_1+c_2)$. As a result, 
the general solution of (\ref{Ermakov-Pinney}) can be determined from a particular solution 
$\rho(t)$ using (\ref{Pinney-oscilador}) and (\ref{oscilador-Pinney}).

The quantum version of the Arnold-Ermakov-Pinney transformation, $\hat E$, 
can be obtained computing the composition of a QAT and an inverse QAT, to give:
%
\begin{align}
  \hat {E}:& \quad {\mathcal H^{(2)}_{t_2}} \;\; \longrightarrow
\quad {\mathcal H^{(1)}_{t_1}} \nonumber\\
    & {\phi}(x_2,t_2)  \longmapsto \;
{\varphi}(x_1,t_1) = 
        \hat E\left( {\phi}(x_2,t_2) \right) \label{AEP} \\
          &\qquad\qquad= E^* \left( \sqrt{b(t_2)}\,e^{-\frac{i}{2}\frac{m}{\hbar}
                 \frac{1}{W_2(t_2)}\frac{\dot b(t_2)}{b(t_2)} {x}^{2}_2}{\phi}(x_2,t_2) \right) \,.\nonumber                
\end{align}

The Quantum Arnold-Ermakov-Pinney transformation is a unitary  map importing solutions of a 
GCK Schr\"odinger equation from solutions of a different, auxiliary GCK Schr\"odinger equation which, 
in particular, can be the one corresponding to a harmonic oscillator. In that case the transformation 
is very similar to the one used in BEC, known 
as {scaling transformation} to transform the time-dependent potential (oscillator traps with  
time-dependent frequencies) into a time-independent harmonic oscillator potential 
\cite{Kagan,Castin}. 
Also, in that case (i.e. for $\dot{f}_2=0,\,W_2=1,\,\dot{f}_1=0,\,W_1=1$) equation (\ref{AEP}) 
reduces to the transformation given by Hartley and Ray \cite{Hartley-Ray}  (this was already given by Lewis and Riesenfeld in \cite{Lewis}).
However, the Quantum Arnold-Ermakov-Pinney transformation allows to choose in a 
suitable way the auxiliary system from which the solutions may be imported.

\section*{Acknowledgments}


Work partially supported by the Fundaci\'on S\'eneca, Spanish MICINN and
Junta de Andaluc\'\i a under projects 08814/PI/08, FIS2011-29813-C02-01
and FQM219-FQM1951, respectively.



\end{document}